\documentclass[prd,amsmath,amssymb,longbibliography,superscriptaddress,twocolumn,nofootinbib,10pt]{revtex4}
\pdfoutput=1
\usepackage{graphicx}
\usepackage{dcolumn}
\usepackage{bm}
\usepackage{amssymb}
\usepackage{latexsym}
\usepackage{booktabs}
\usepackage{amsmath}
\usepackage{multirow}
\usepackage[colorlinks=true, linkcolor=blue, citecolor=blue]{hyperref}
\usepackage{subcaption}
\usepackage[utf8]{inputenc}
\usepackage{xcolor}

\newcommand{\be}{\begin{equation}}
\newcommand{\ee}{\end{equation}}
\newcommand{\bea}{\begin{eqnarray}}
\newcommand{\eea}{\end{eqnarray}}

\begin{document}

\title{Model-Independent Reconstruction of Quintessence Potential and Kinetic Energy from DESI DR2 and Pantheon+ Supernovae}

\author{Shengjia Wang}
\affiliation{School of Physics and Optoelectronic Engineering, Yangtze University, Jingzhou 434023, China}

\author{Tian-Nuo Li}
\affiliation{Liaoning Key Laboratory of Cosmology and Astrophysics, College of Sciences, Northeastern University, Shenyang 110819, China}

\author{Tonghua Liu}
\email{liutongh@yangtzeu.edu.cn}
\affiliation{School of Physics and Optoelectronic Engineering, Yangtze University, Jingzhou 434023, China}

\author{Guo-Hong Du}
\affiliation{Liaoning Key Laboratory of Cosmology and Astrophysics, College of Sciences, Northeastern University, Shenyang 110819, China}

\begin{abstract}
We present a  model-independent reconstruction of the quintessence scalar field's dynamics-both its potential and kinetic energy-directly from the latest cosmological observations. Our analysis combines DESI DR2 baryon acoustic oscillation measurements with the Pantheon+ Type Ia supernova compilation, employing Gaussian process  with four distinct covariance kernels to avoid theoretical priors on the potential's functional form. Key findings reveal a monotonically decreasing potential with redshift, consistent with thawing quintessence, and a kinetic energy that crosses zero near $z \sim 1$, marking the dark energy–matter equality epoch. Notably, while apparent negative kinetic energy values emerge at intermediate redshifts ($0.5<z<1.0$), these are statistical artifacts within uncertainties, arising from error amplification in derivative reconstruction rather than new physics. Our results demonstrate the power of non-parametric methods to constrain dynamical dark energy and show minimal dependence on the choice of cosmological priors, whether from local (SH0ES) or early-universe (Planck) measurements.
\end{abstract}

\maketitle
\section{\label{Introduction}INTRODUCTION}

The discovery of the Universe's late-time acceleration represents a foundational result in modern cosmology. Initial evidence emerged from SN Ia observations by the Supernova Cosmology Project \cite{1999AJ....117..707R} and the High-z Supernova Search Team \cite{1998AJ....116.1009R}, which revealed a systematic dimming of high-redshift sources. The prevailing theoretical explanation invokes dark energy, with the $\Lambda$ Cold Dark Matter ($\Lambda$CDM) model constituting the standard cosmological paradigm. The robustness of this model is underpinned by its consistency with multiple independent probes, including precision measurements of the cosmic microwave background (CMB) anisotropies by the Planck satellite \cite{2016A&A...594A..14P,2020A&A...641A...6P}, large-scale structure statistics from galaxy surveys such as the Sloan Digital Sky Survey (SDSS) \cite{2015PhRvD..92l3516A, 2017MNRAS.470.2617A, 2021PhRvD.103h3533A}, and the distance-ladder calibration provided by type Ia supernova (SN Ia) samples from dedicated campaigns \cite{2014A&A...568A..22B, 2018ApJ...859..101S, 2022ApJ...938..110B} and the Hubble Space Telescope \cite{1997eds..proc....6T, 2009AJ....138.1271D, 2016ApJ...826...56R}.

The $\Lambda$CDM model faces significant challenges, most notably the Hubble tension-a statistically significant discrepancy between the Hubble constant ($H_0$) inferred from Planck CMB data ($H_0 = 67.4 \pm 0.5~\mathrm{km~s^{-1}~Mpc^{-1}}$) \cite{2020A&A...641A...6P} and that measured by the SH0ES collaboration using the cosmic distance ladder ($H_0 = 73.04 \pm 1.04~\mathrm{km~s^{-1}~Mpc^{-1}}$) \cite{2022ApJ...934L...7R}. Further pressure comes from recent baryon acoustic oscillation (BAO) measurements by Dark Energy Spectroscopic Instrument (DESI). Combined analyses of its first data release (DR1) and DR2 with CMB data reveal a $3.1\sigma$ preference for dynamical dark energy over a cosmological constant \cite{2025PhRvD.111b3532L,2025PhRvD.112h3515A}. This preference persists when DESI BAO are jointly analyzed with supernova datasets such as PantheonPlus, Union3, or DES-Y5, yielding significances between $2.8\sigma$ and $4.2\sigma$ for the Chevallier–Polarski–Linder (CPL) parametrization. While CPL offers a useful phenomenological description, its limited theoretical motivation underscores the need for physically grounded alternatives. In light of these persistent tensions, the pursuit of well-motivated dynamical dark energy frameworks has become both timely and necessary.

The challenges faced by current theories are driving researchers to explore new frameworks beyond the standard cosmological model. These efforts have primarily bifurcated into two avenues: one stems from modifications to the geometric nature of gravity, such as extending Einstein's theory of gravitation within frameworks like $f(R)$ \cite{2012PhR...513....1C}, $f(T)$ \cite{2016RPPh...79j6901C} and $f(R,T)$ \cite{2011PhRvD..84b4020H}; the other is grounded in matter fields, seeking to introduce a dynamical form of dark energy into the cosmic energy budget. Among the various dynamical dark energy models, scalar field theories have become the most extensively studied and compelling candidates, owing to their theoretical self-consistency and rich phenomenological implications \cite{2024JHEP...05..327Y,2024PhLB..85538826S,2024JCAP...11..025N,2024PhRvD.110h3528W,2025PhRvD.112f3508S,2025arXiv250319428A,2025PhRvD.112f3551G,2026arXiv260107361L}.
A recent study has employed redshift binning and polynomial interpolation to reconstruct the equation of state of dark energy. After incorporating DESI data, the result indicates that current observational data favor a dynamical dark energy component over a static cosmological constant \cite{2025arXiv250622953L}. Moreover, research extends to quintessence models, phantom fields, Quintom models (enabling $w = -1$ crossing), massive vector fields, holographic dark energy \cite{2025JCAP...07..047L,2024PhRvD.110f3503T,2025PDU....4701747A,2025EPJC...85..608L,2025Univ...11...85H}, interacting dark energy \cite{2024PhRvL.133y1003G,2025PDU....4801848M,2024ApJ...976....1L,2025SCPMA..6910413L,2025EPJC...85..224Y,2025arXiv250315659Z,2025arXiv250400994P,2025PhRvD.111j3534Y}, and early dark energy scenarios \cite{2025PhRvD.112f3548C,2025PhRvD.111l3507Q,2024PhRvD.110h3501S}. Comprehensive reviews of these developments are available in \cite{2012IJMPD..2130002Y,2017AdSpR..60..166A,2018PhR...775....1B,2016arXiv161200345W,2023CoTPh..75k7401W,2025arXiv250524732C}.

In the study of dark energy properties, research methodologies have evolved from strongly parameter-dependent assumptions toward more flexible, model-independent reconstructions. Early work typically adopted specific scalar potential forms, such as the linear potential \cite{1988PhRvD..37.3406R} or tracker potential \cite{1999PhRvL..82..896Z}, which inherently constrained the possible dynamical behaviors of dark energy. To overcome this limitation, recent efforts have shifted to non-parametric reconstruction techniques, among which Gaussian process (GP)  has emerged as a powerful model-independent tool for inferring the morphology of the scalar potential directly from observational data. In one representative application, \citet{2023A&A...673A...9S} applied GP directly to the cosmic expansion history $H(z)$ and derived the scalar potential through the dynamical equations of quintessence. Their reconstruction, based on DESI DR1, SDSS, and 30 new cosmic chronometer data points, reveals a monotonically decreasing potential at redshifts $z < 1$, qualitatively consistent with thawing-type models. In a related methodology,  \citet{2025arXiv251102220Z} employed a GP-based reconstruction of the dimensionless luminosity distance $D(z)$ and its derivatives, from which they inferred the cosmological evolution of key parameters including the deceleration parameter $q(z)$ and the dark energy equation of state $w(z)$. These approaches exemplify the growing emphasis on data-driven, assumption-minimal frameworks in probing dark energy dynamics.

Inspired by the aforementioned studies, we introduce a non-parametric reconstruction based on GP that provides a model-independent approach to reconstruct the dynamics of the quintessence field directly from combinations of recent DESI BAO DR2 and SN Ia datasets. Unlike traditional approaches that presuppose a specific potential $V(\phi)$-thereby introducing theoretical bias-our method performs a non-parametric, model-independent reconstruction of the scalar field's potential and kinetic energy without imposing any functional form assumptions.

The paper is organized as follows: We present the derivation of the scalar field potential for dark energy from the Friedmann equations in Section \ref{dyn}. Section \ref{datamet} describes the observational datasets and GP methodology employed in this work. The results of this study are presented in Section \ref{results}. Finally, Section \ref{conc} offers conclusions and discussions of our findings.

\section{\label{dyn}Cosmological Dynamics}

The Friedmann equations, derived from Einstein's field equations under the assumption of a homogeneous and isotropic universe, provide the fundamental framework for describing cosmic expansion dynamics. For a quintessence cosmological model in a spatially flat universe ($k=0$) and neglecting radiation density, the Friedmann equation and acceleration equation are given by:
\begin{align}
H^2 &= \frac{8\pi G}{3}(\rho_m + \rho_\phi), \label{EF_flat} \\
\dot{H}+H^2 &= -\frac{4\pi G}{3} (\rho_m + \rho_\phi + 3p_\phi). \label{AC_flat}
\end{align}
where $\rho_{m}$ is the matter density, the scalar field $\phi$ has energy density $\rho_{\phi} = \frac{1}{2}\dot{\phi}^2+V(\phi)$ and pressure $p_{\phi} = \frac{1}{2}\dot{\phi}^2-V(\phi)$.

Rewriting Eqs. \eqref{EF_flat} and \eqref{AC_flat} explicitly in terms of the scalar field components yields:
\begin{align}
H^2 &= \frac{8\pi G}{3}\left[\rho_m +\frac{\dot\phi^2}{2}+V(\phi)\right], \label{EF2_flat} \\
\dot{H}+H^2 &= -\frac{4\pi G}{3}\left[\rho_m +2\dot\phi^2-2V(\phi)\right]. \label{EA2_flat}
\end{align}
Eliminating $\dot{\phi}^2$ from these equations allows us to solve for the scalar field potential:
\begin{equation}
V(\phi) = \frac{3H^2+\dot H}{8\pi G}-\frac{\rho_m}{2}. \label{EP_flat}
\end{equation}

For numerical convenience, we define the dimensionless potential $U(\phi)\equiv \kappa V(\phi)$ with $\kappa=8\pi G/3H_0^2$. Using the redshift-space transformation $d/dt=-H(1+z)d/dz$, the dimensionless potential becomes:
\begin{equation}
U(z)=E^2-\frac{E(1+z)}{3}\frac{dE}{dz}-\frac{\Omega_{m0}(1+z)^3}{2}, \label{UphiEz_flat}
\end{equation}
where $E(z)\equiv H(z)/H_0$, $\Omega_{m0}\equiv\kappa\rho_{m0}$ (the present-day matter density parameter), and we adopt pressureless matter conservation: $\kappa\rho_m=\Omega_{m0}(1+z)^3$.

Similarly, eliminating $V(\phi)$ gives the kinetic energy $T \equiv \dot{\phi}^2/2$, whose dimensionless form $\tau \equiv \kappa T$ is:
\begin{equation}
\tau(z) = \frac{E(1+z)}{3}\frac{dE}{dz} - \frac{\Omega_{m0}(1+z)^3}{2}. \label{tauz_flat}
\end{equation}
Both $U(z)$ and $\tau(z)$ are expressed in terms of $E(z)$ and its derivative, enabling direct reconstruction from cosmological expansion data.

For distance-based reconstruction, we relate these quantities to the dimensionless transverse comoving distance $D_M(z)$. In a flat universe, $D_M(z) = D_C(z)$ (line-of-sight comoving distance), which satisfies $D_C'(z) = 1/E(z)$. This leads to $E(z) = 1/D_M'(z)$ and $\frac{dE}{dz} = -D_M''(z)/[D_M'(z)]^2$. Substituting these into Eqs. \eqref{UphiEz_flat} and \eqref{tauz_flat} produces the final expressions:
\begin{align}
U(z) = \frac{1}{[D_M'(z)]^2} + \frac{(1+z)D_M''(z)}{3[D_M'(z)]^3} - \frac{\Omega_{m0}(1+z)^3}{2}, \label{UzDC_flat}
\end{align}
\begin{align}
\tau(z) = -\frac{(1+z)D_M''(z)}{3[D_M'(z)]^3} - \frac{\Omega_{m0}(1+z)^3}{2}. \label{tau_final}
\end{align}
Here, primes denote differentiation with respect to redshift $z$. This formulation enables direct reconstruction of the scalar field potential and kinetic energy from $D_M(z)$ constrained by distance indicators such as supernovae and BAO.

\section{\label{datamet}Dataset and Methodology}

We utilize PantheonPlus+SH0ES as the primary observational dataset for reconstruction, complemented by DESI DR2 BAO data to enhance the precision of GP reconstruction. We compare these results with the $\Lambda$CDM model and test the sensitivity of our GP approach to cosmological priors by applying two distinct $H_0$ priors:
\begin{enumerate}
    \item \textbf{Local distance-ladder prior}: Derived from PantheonPlus+SH0ES within $\Lambda$CDM \cite{2022ApJ...938..110B}, with $H_0 = 73.6 \pm 1.1$ km s$^{-1}$ Mpc$^{-1}$, $\Omega_{m0} = 0.334 \pm 0.018$, $M_B = -19.25 \pm 0.03$ mag, and corresponding sound horizon $r_d = 135.8 \pm 1.3$ Mpc.
    \item \textbf{Early-universe prior}: Based on Planck 2018 CMB constraints \cite{2020A&A...641A...6P}, with $H_0 = 68.17 \pm 0.28$ km s$^{-1}$ Mpc$^{-1}$, $\Omega_{m0} = 0.315 \pm 0.007$, directly measured $r_d = 147.09 \pm 0.26$ Mpc, and $M_B = -19.40 \pm 0.01$ mag (from Planck + late-time probes).
\end{enumerate}
Note that all cosmological parameters of these two priors are assumed to follow a Gaussian distribution. This choice is consistent with the characteristics of the random errors inherent in the observational data, satisfies the mathematical requirements of the GP methodology, and minimizes potential model bias introduced by the prior. This approach enables a rigorous investigation of the prior dependence in our reconstruction and facilitates an assessment of its cosmological implications.

\subsection{Observational Data}

\textbf{BAO}: We use DESI DR2 \cite{2025PhRvD.112h3515A}, the most comprehensive BAO dataset to date, which combines Luminous Red Galaxies (LRG), Emission Line Galaxies (ELG), Quasars (QSO), and the Lyman-$\alpha$ forest (Ly$\alpha$). DESI DR2 provides precise measurements of $D_M/r_d$ (the transverse comoving distance normalized by the sound horizon $r_d$) and $D_H/r_d$ (the Hubble distance normalized by the sound horizon $r_d$). To reconstruct the physical $D_M(z)$, we multiply $D_M/r_d$ by the $r_d$ value from each prior. To fully utilize the BAO information, we also convert the $D_H/r_d$ measurements into constraints on $D_M$. In a flat cosmological model, there exists a theoretical relationship between the transverse comoving distance and the radial distance. We leverage this relationship to convert $D_H$ into $D_M$ through numerical integration, thereby obtaining additional distance constraints. The complete BAO dataset is presented in Table \ref{tab:desi_dm_rd}.

\textbf{SN Ia}: We utilize the PantheonPlus+SH0ES compilation, which includes 1,701 high-quality light curves spanning $0.01 < z < 2.3$. We apply conservative selection criteria: excluding low-redshift ($z < 0.01$) events to minimize peculiar velocity and host galaxy biases, removing known calibrator supernovae to avoid circularity, and retaining 1,550 high-quality SN Ia with corrected $m_B$ magnitudes.

We account for statistical and systematic uncertainties using Monte Carlo error propagation with Cholesky decomposition of the full covariance matrix. We convert the distance modulus $\mu(z) = m_B(z) - M_B$ (where $m_B(z)$ is the observed apparent magnitude and $M_B$ is the absolute magnitude) to $D_M(z)$ via:
\begin{equation}
D_M(z) = \frac{c}{H_0} \frac{1}{1+z} \left(10^{\mu(z)/5 - 5}\right).
\end{equation}
The combination of these datasets enables robust reconstruction of the cosmological expansion history and scalar field dynamics.
\begin{table}
    \centering
    \caption{Effective redshifts and measured $D_M / r_d$ and $D_H / r_d$ values (with 1$\sigma$ uncertainties) for DESI DR2 BAO tracers \cite{2025PhRvD.112h3515A}.}
    \label{tab:desi_dm_rd}
    \setlength{\tabcolsep}{8pt}
    \renewcommand{\arraystretch}{1.2}
    \begin{tabular}{lccc}
        \hline
        \hline
        Tracer & $z_{\mathrm{eff}}$ & $D_M/r_d$ & $D_H/r_d$ \\
        \hline
        LRG1 & 0.510 & $13.588 \pm 0.167$ & $21.863 \pm 0.425$ \\
        LRG2 & 0.706 & $17.351 \pm 0.177$ & $19.455 \pm 0.330$ \\
        LRG3 & 0.922 & $17.577 \pm 0.178$ & $14.176 \pm 0.221$ \\
        ELG1 & 0.955 & $17.803 \pm 0.335$ & $12.817 \pm 0.516$ \\
        ELG2 & 1.321 & $27.601 \pm 0.318$ & $8.632 \pm 0.101$ \\
        QSO & 1.484 & $30.512 \pm 0.760$ & $17.577 \pm 0.213$ \\
        Ly$\alpha$ & 2.330 & $38.988 \pm 0.531$ & $17.5803 \pm 0.297$ \\
        \hline
        \hline
    \end{tabular}
\end{table}

\subsection{Gaussian Process Method}
GP is a model-independent, non-parametric technique \cite{2012JCAP...06..036S},which is widely used in cosmology \cite{2012PhRvD..85l3530S,2020JCAP...04..053J,2021EPJC...81..903L,2018PhRvD..97l3501J,2025ApJ...981L..24L}. It characterizes the distribution of the function without imposing a predefined form, and the final reconstruction result is not strictly limited by the selection of the mean function and the kernel. The hyperparameters of the GP are optimized to capture the deviation from the mean value, and the selection of the kernel is also adjusted so that the data itself can determine the final form of the distance-redshift relationship. This method has been widely recognized in the literature~\cite{2012PhRvD..85l3530S,2013PhRvD..87b3520S,2017JCAP...09..031A} for achieving a largely data-driven reconstruction.

For observational data $\mathcal{D} = \{(x_i, y_i)\}$, the GP framework assumes $\mathbf{y} \sim \mathcal{N} \left( \boldsymbol{\mu}, \mathbf{K}_{\mathbf{xx}} + \mathbf{C} \right)$, where $\boldsymbol{\mu}$ is the mean function (set to zero), $\mathbf{C}$ is the observational uncertainty matrix, and $\mathbf{K}_{\mathbf{xx}} = K(\mathbf{x}, \mathbf{x})$ is the covariance matrix defined by a kernel function. The posterior distribution of the target function $f_* = f(\mathbf{x}_*)$ at reconstruction points $\mathbf{x}_*$ is:

\begin{equation}
\begin{aligned}
f_* | \mathbf{y}, \mathbf{x}, \mathbf{x}_*
\sim \mathcal{N}\Bigl(
    &\mathbf{K}_{* \mathbf{x}} (\mathbf{K}_{\mathbf{xx}} + \mathbf{C})^{-1} \mathbf{y}, \\
    &\mathbf{K}_{**} - \mathbf{K}_{* \mathbf{x}} (\mathbf{K}_{\mathbf{xx}} + \mathbf{C})^{-1} \mathbf{K}_{\mathbf{x} *} \Bigr),
\end{aligned}
\end{equation}
where $\mathbf{K}_{* \mathbf{x}} = K(\mathbf{x}_*, \mathbf{x})$, $\mathbf{K}_{\mathbf{x} *} = K(\mathbf{x}, \mathbf{x}_*)$, and $\mathbf{K}_{**} = K(\mathbf{x}_*, \mathbf{x}_*)$.

Kernel selection is crucial for the GP reconstruction, because even with the same data, different kernels will yield different results. Therefore, we employed four widely used covariance functions and separately examined the reliability of their reconstructions. The four variance functions are as follows:

\textbf{RBF kernel} (Radial basis function):
\begin{equation}
k_{\mathrm{RBF}}(r) = \sigma_f^2 \exp\bigl(-r^2/(2\ell^2)\bigr),
\end{equation}
where $r=|x_i-x_j|$, $\sigma_f^2$ is the signal variance, and $\ell$ is the length scale. This infinitely differentiable kernel generates smooth functions.

\textbf{Matérn-5/2 kernel}:
\begin{equation}
k_{\mathrm{M52}}(r) = \sigma_f^2 \Bigl(1 + \frac{\sqrt{5}r}{\ell} + \frac{5r^2}{3\ell^2}\Bigr)
    \exp\!\Bigl(-\frac{\sqrt{5}r}{\ell}\Bigr),
\end{equation}
yielding twice-differentiable functions that balance smoothness and flexibility.

\textbf{Matérn-7/2 kernel}:
\begin{equation}
\begin{aligned}
k_{\mathrm{M72}}(r) = \sigma_f^2 \Bigl(1 &+ \frac{\sqrt{7}r}{\ell} + \frac{14r^2}{5\ell^2} \\
    &+ \frac{7\sqrt{7}r^3}{15\ell^3}\Bigr) \exp\!\Bigl(-\frac{\sqrt{7}r}{\ell}\Bigr),
\end{aligned}
\end{equation}
producing three-times-differentiable functions with enhanced smoothness.

\textbf{Matérn-9/2 kernel}:
\begin{equation}
\begin{aligned}
k_{\mathrm{M92}}(r) = \sigma_f^2 \Bigl(1 &+ \frac{3r}{\ell} + \frac{27r^2}{7\ell^2} \\
    &+ \frac{18r^3}{7\ell^3} + \frac{27r^4}{35\ell^4}\Bigr) \exp\!\Bigl(-\frac{3r}{\ell}\Bigr),
\end{aligned}
\end{equation}
the smoothest option (four-times differentiable) among the Matérn class considered.

All hyperparameters ($\sigma_f, \ell$) are marginalized via Markov Chain Monte Carlo (MCMC) sampling for each kernel, enabling a fully Bayesian treatment that incorporates their uncertainties.

\begin{figure}[h]
    \includegraphics[width=0.48\textwidth]{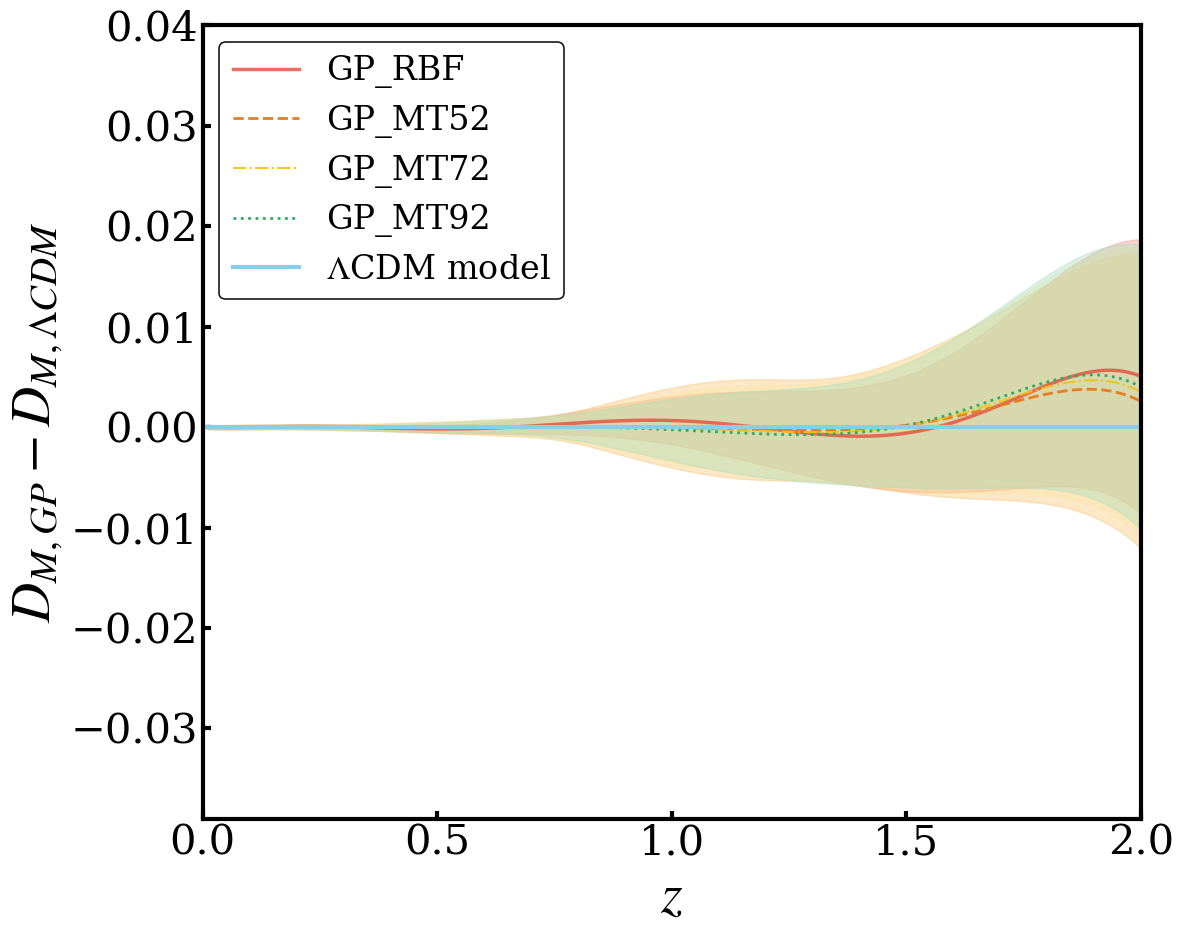}
    \caption{Reliability assessment of reconstruction using four GP kernels based on Pantheon+SH0ES data ($H_0=67.3$, $\Omega_{m0}=0.3$).}
    \label{fig:sample}
\end{figure}

 In the verification of the reliability of the four kernel function reconstructions, we conducted a numerical simulation using the flat $\Lambda$CDM model to obtain a curve depicting the evolution of the $D_M$ with redshift $z$. Then, based on this model, we first fitted the distance-redshift relationship, and used the ``Pantheon+'' observational error generation to generate simulated data points and their related errors that were consistent with the theoretical curve. Then, we applied GP with four different kernel functions to reconstruct the cosmological relationship, thereby obtaining four reconstructed curves and their corresponding uncertainty ranges, as shown in Fig.~\ref{fig:sample}. Systematic comparisons showed that the residuals between the reconstructed curves and the theoretical curves all fell within the 1$\sigma$ confidence interval, indicating that each kernel function can provide reliable reconstructions of the redshift-distance relationship, meeting the precision requirements of cosmological analysis. It is noteworthy that the reconstructed $D_M(z)$ curves exhibit an oscillatory behavior around the fiducial $\Lambda$CDM result, and this feature becomes more pronounced at the high-redshift end ($z > 1.0$). This oscillatory characteristic is closely related to the subsequent analysis of the physical validity of the quintessence potential and the evolution of kinetic energy: the small-scale fluctuations in the reconstructed $D_M(z)$ (particularly the oscillations at high redshifts) are amplified when deriving higher-order derivatives (e.g., the second derivative of $D_M(z)$), which is one of the key factors leading to statistical artifacts in the kinetic energy reconstruction at intermediate and high redshifts, such as the appearance of unphysical negative $\tau(z)$ values.

\begin{figure*}[htbp]
    \centering
    \begin{tabular}{c c c c}
        & \textbf{Comoving distance $D_M(z)$} & \textbf{$D_M'(z)$: First derivative} & \textbf{$D_M''(z)$: Second derivative} \\

        \multirow{2}{*}[4.5cm]{\rotatebox[origin=c]{90}{\textbf{Planck 2018 prior}}} &
        \begin{subfigure}{0.32\textwidth}
            \centering
            \includegraphics[width=\linewidth]{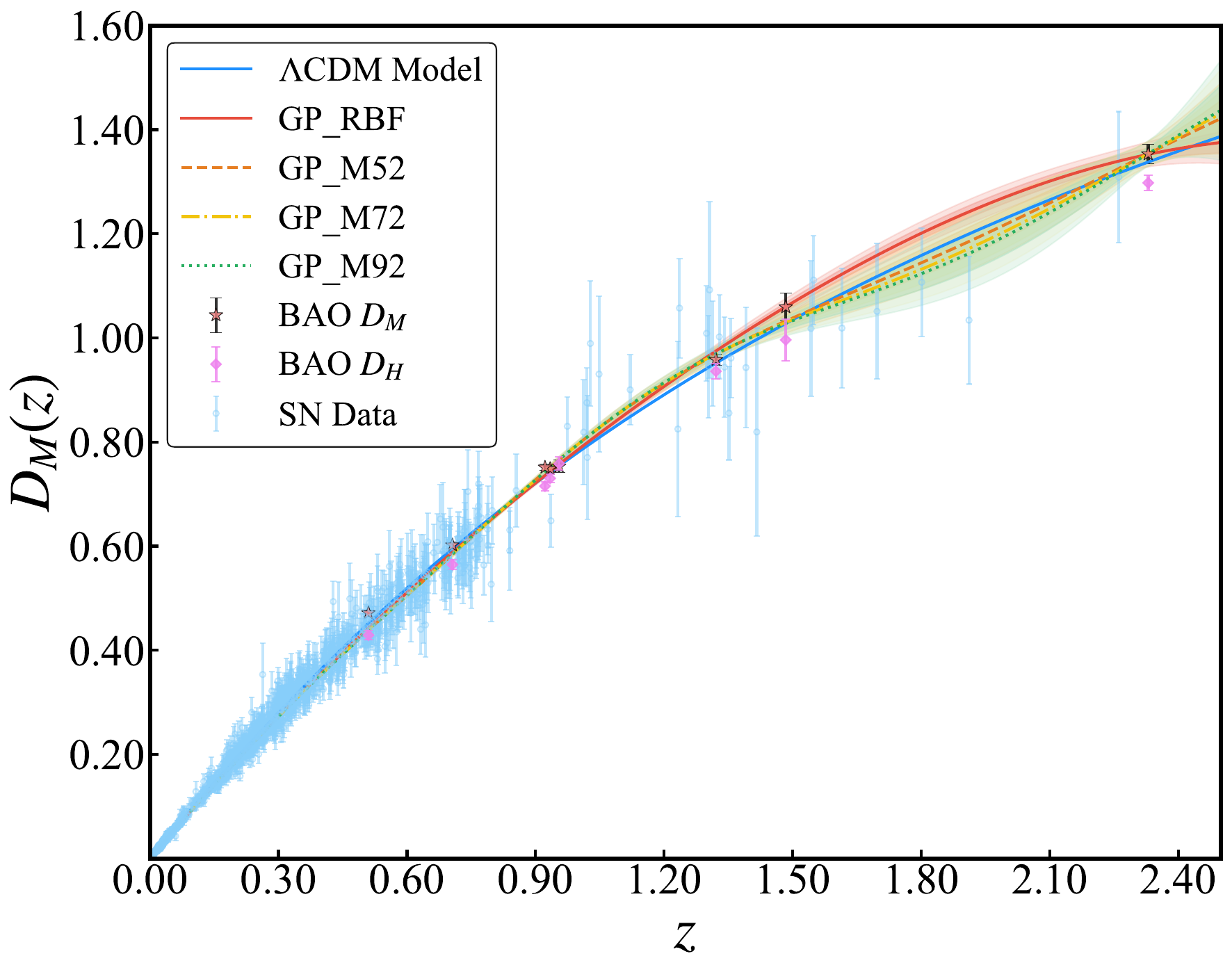}
            \caption{}
            \label{fig:planck_dc}
        \end{subfigure} &
        \begin{subfigure}{0.32\textwidth}
            \centering
            \includegraphics[width=\linewidth]{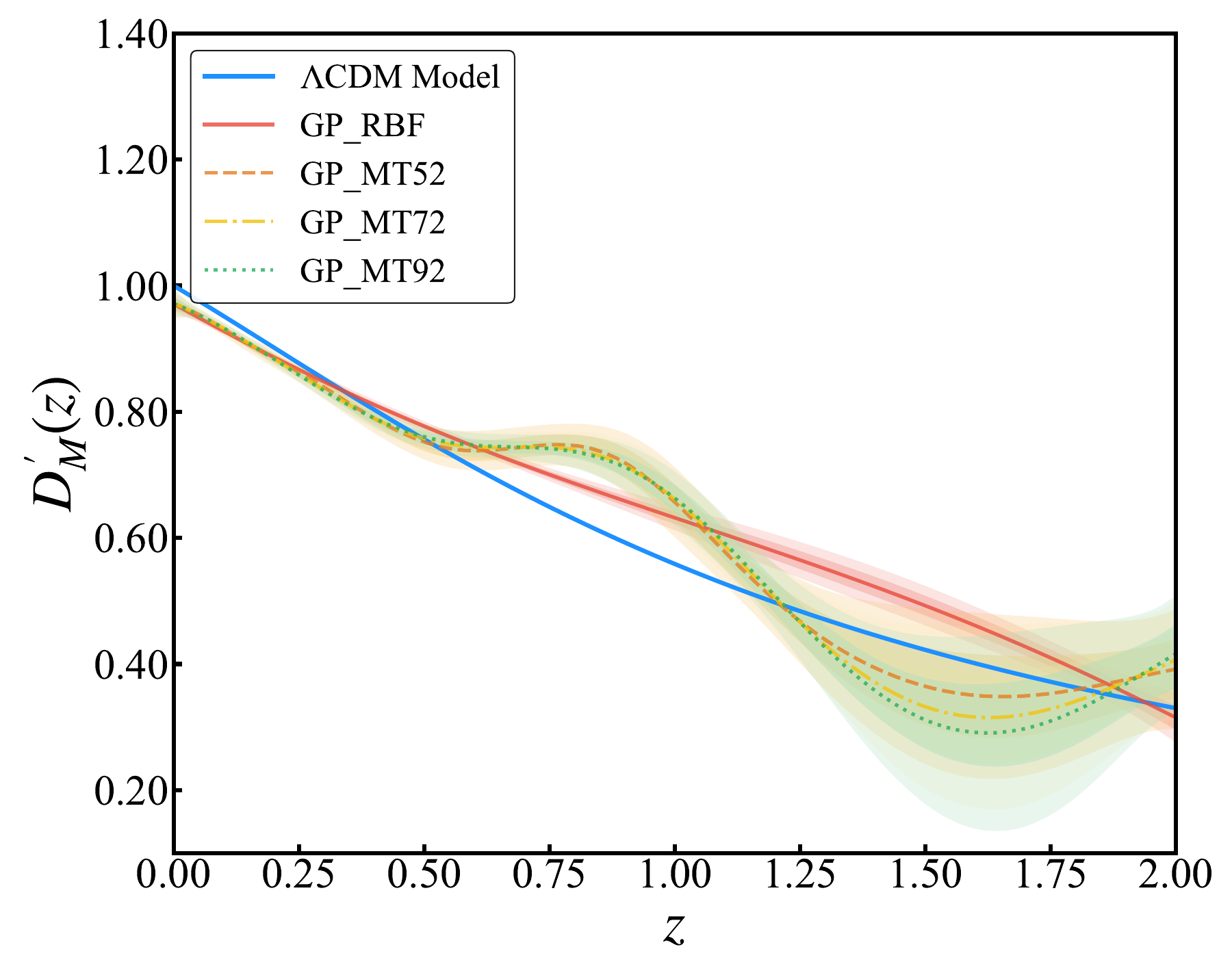}
            \caption{}
            \label{fig:planck_dc1}
        \end{subfigure} &
        \begin{subfigure}{0.32\textwidth}
            \centering
            \includegraphics[width=\linewidth]{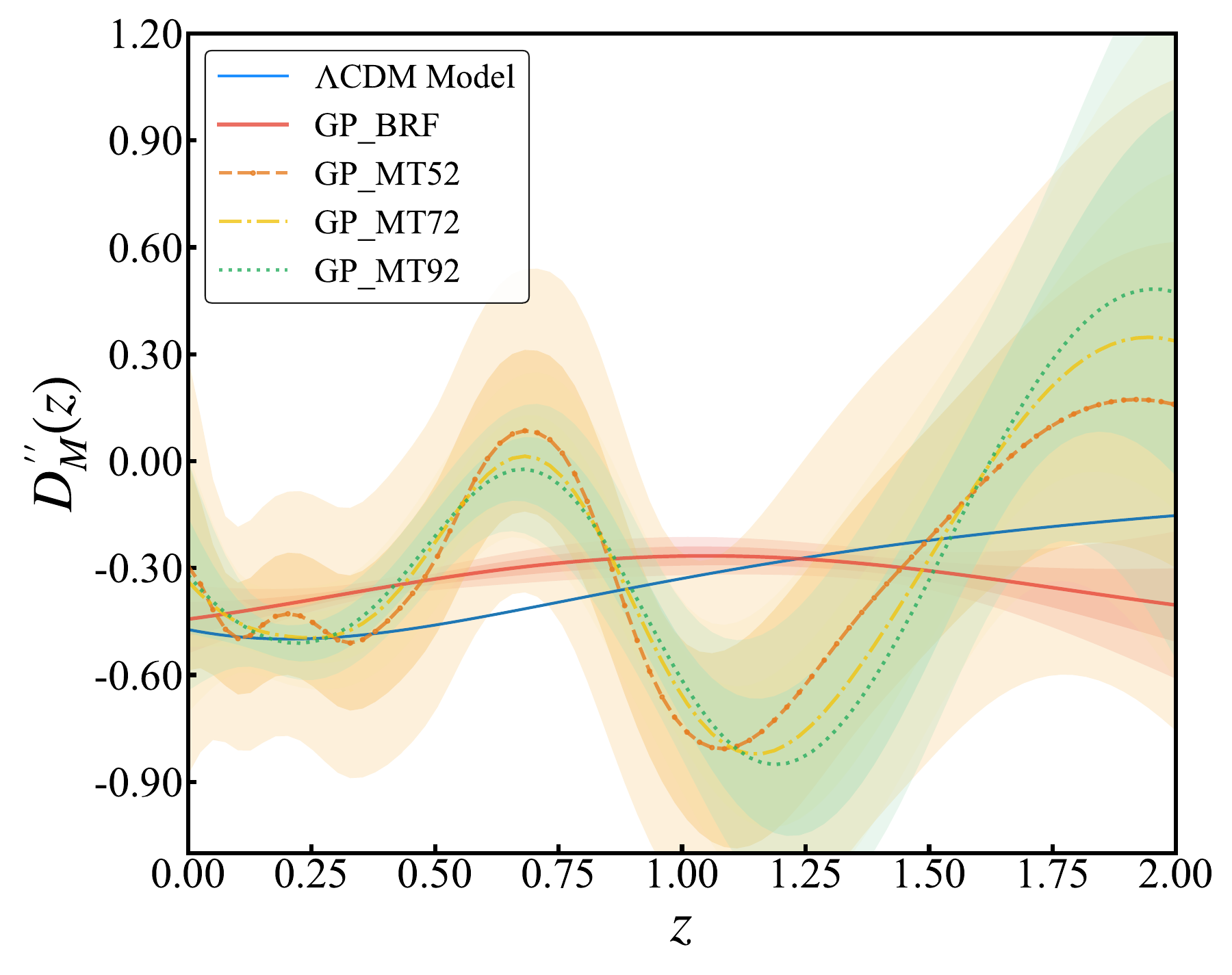}
            \caption{}
            \label{fig:planck_dc2}
        \end{subfigure} \\

        & & & \\

        \multirow{2}{*}[5cm]{\rotatebox[origin=c]{90}{\textbf{PantheonPlus+SH0ES prior}}} &
        \begin{subfigure}{0.32\textwidth}
            \centering
            \includegraphics[width=\linewidth]{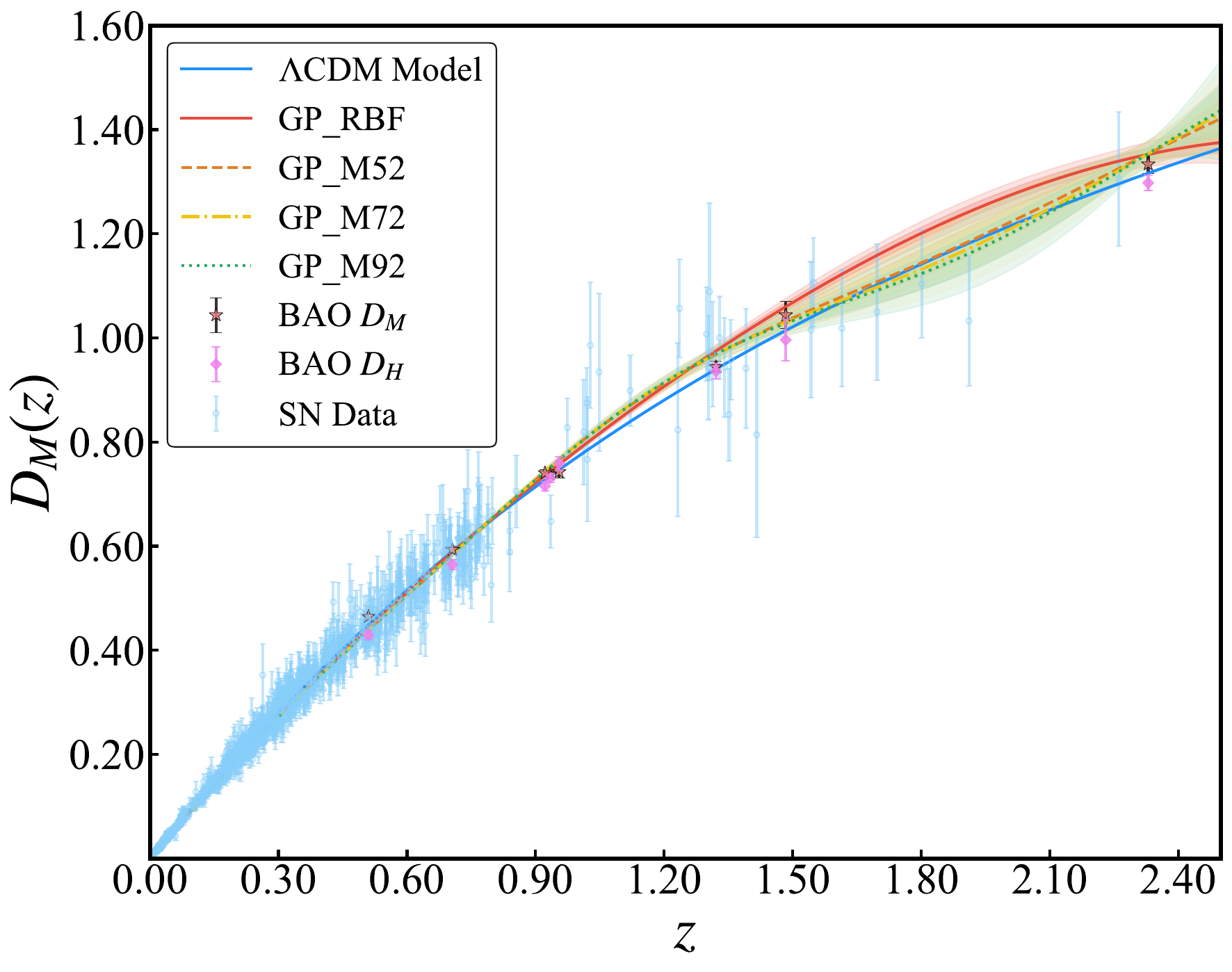}
            \caption{}
            \label{fig:pantheon_dc}
        \end{subfigure} &
        \begin{subfigure}{0.32\textwidth}
            \centering
            \includegraphics[width=\linewidth]{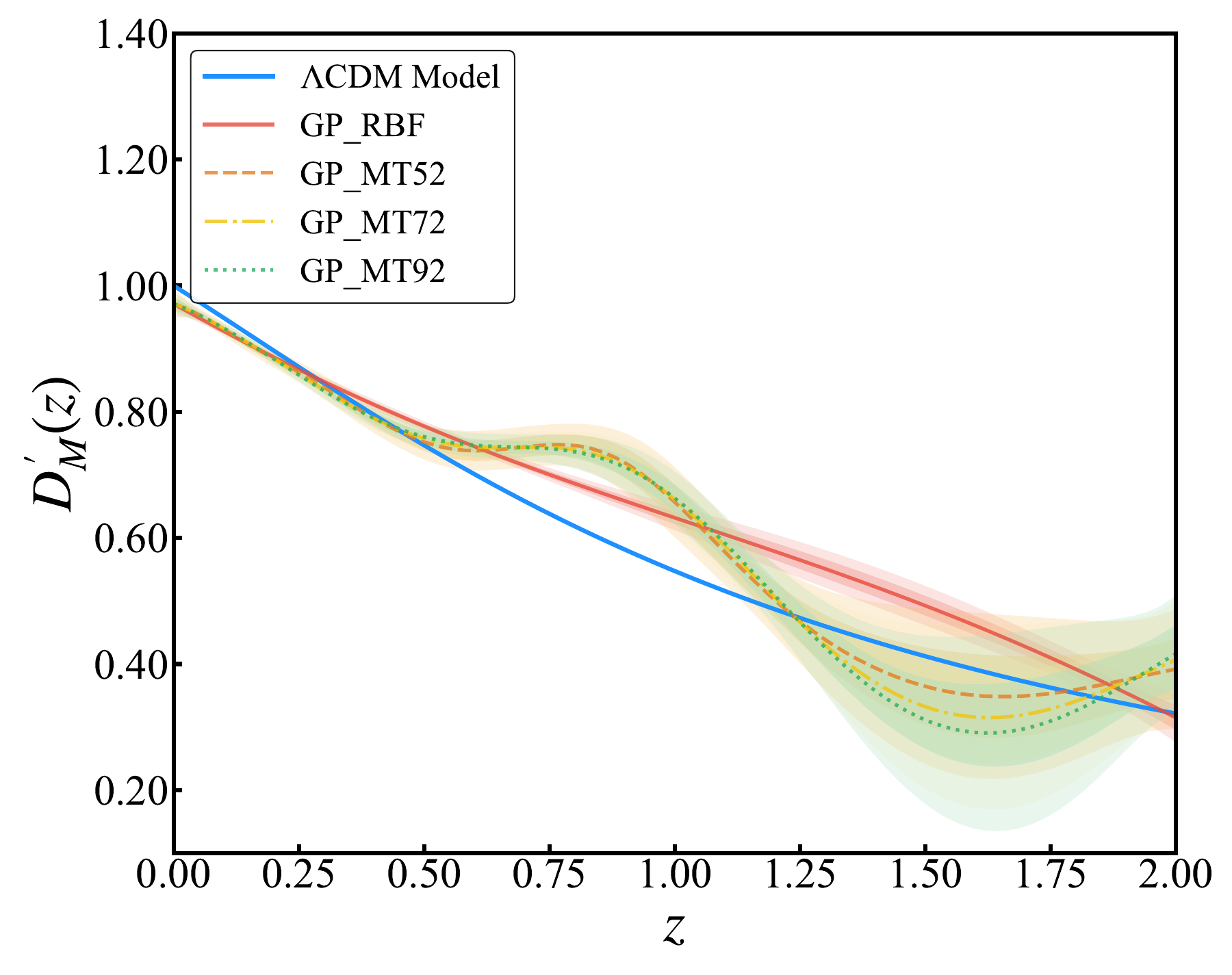}
            \caption{}
            \label{fig:pantheon_dc1}
        \end{subfigure} &
        \begin{subfigure}{0.32\textwidth}
            \centering
            \includegraphics[width=\linewidth]{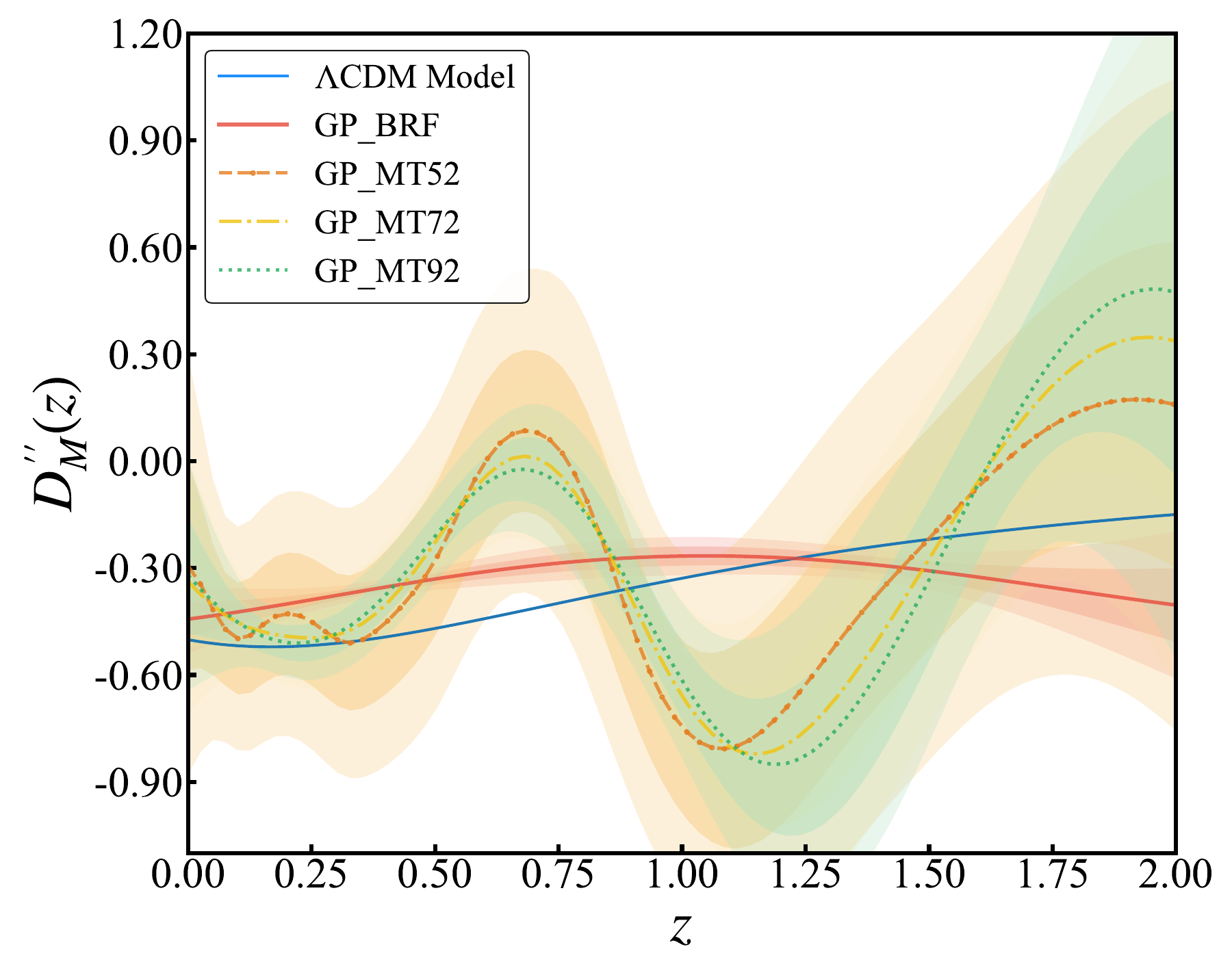}
            \caption{}
            \label{fig:pantheon_dc2}
        \end{subfigure} \\
    \end{tabular}

    \caption{Non-parametric reconstruction of comoving distance $D_M(z)$ and its derivatives from SN Ia and BAO data using four kernel functions. \textbf{Top row}: Planck~2018 prior. \textbf{Bottom row}: PantheonPlus+SH0ES prior.  Solid lines represent mean reconstructions from different kernels, while shaded regions denote 1$\sigma$ and $2\sigma$ confidence regions.}
    \label{fig:comoving_reconstruction}
\end{figure*}
We implement the reconstruction using GaPP \cite{2012JCAP...06..036S} and developed parallelized code to sample the multivariate Gaussian distributions of $D_M(z)$ and its derivatives. This allows robust reconstruction of $U(z)$ and $\tau(z)$ across all kernel choices. The complete reconstructions of $D_M(z)$ and its derivatives are presented in Fig.~\ref{fig:comoving_reconstruction}.

To contextualize our results, we compare the reconstructed $U(z)$ with two canonical scalar field dark energy models, and the curves of the two potentials shown in Fig.~3 are their best-fit lines to the combined observational data (DESI DR2 BAO + Pantheon+SH0ES SN Ia). The best-fit parameter values of the two potentials, obtained via Markov Chain Monte Carlo (MCMC) sampling with the same cosmological priors (Planck 2018 and PantheonPlus+SH0ES) adopted in our $U(z)$ reconstruction, are listed as follows:

1). Quadratic free-field potential \cite{1991PhRvD..44..352R,2009IJMPD..18..621U}:
\begin{equation}
U_{\mathrm{FF}}(\Phi) = \frac{1}{2} \mu^2 \Phi^2,
\end{equation}
where $\mu\equiv m/H_0$ (scalar field mass normalized by $H_0$) and $\Phi\equiv\sqrt{8\pi G/3}\,\phi$ (dimensionless scalar field). The best-fit value is $\mu = 0.012 \pm 0.003$ (consistent for both Planck 2018 and PantheonPlus+SH0ES priors, with $\Delta\mu < 0.001$).

2). Peebles-Ratra power-law potential \cite{1988ApJ...325L..17P,1988PhRvD..37.3406R}:
\begin{equation}
U_{\mathrm{PL}}(\Phi) = \frac{1}{2} \beta \Phi^{-\alpha},
\end{equation}
with constant $\beta = \frac{8(\alpha+4)}{3(\alpha+2)} \left[ \frac{2}{3}\alpha(\alpha+2)\right]^{\alpha/2}$.
For $\alpha = 0$, this reduces to the $\Lambda$CDM scenario. The best-fit power-law index is $\alpha = 0.86 \pm 0.15$ (Planck 2018 prior) and $\alpha = 0.91 \pm 0.17$ (PantheonPlus+SH0ES prior), with the normalization constant $\beta$ determined by the best-fit $\alpha$ for each prior.

\section{\label{results}Results}

\begin{figure*}[htbp]
    \centering
    \begin{subfigure}[t]{0.48\textwidth}
        \centering
        \includegraphics[width=\textwidth]{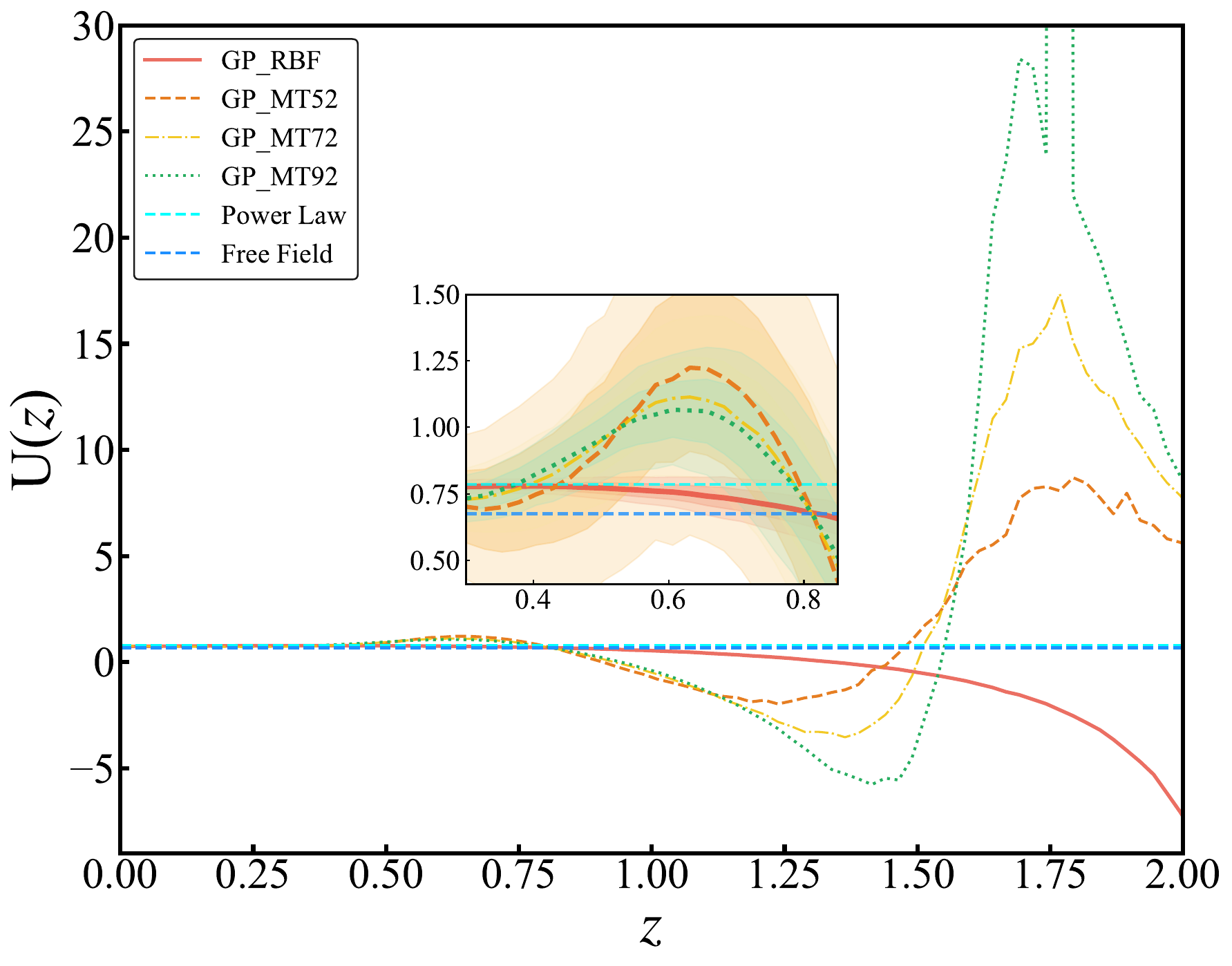}
        \caption{$U(z)$ reconstruction (Planck 2018 prior)}
        \label{fig:u_planck}
    \end{subfigure}
    \hfill
    \begin{subfigure}[t]{0.48\textwidth}
        \centering
        \includegraphics[width=\linewidth]{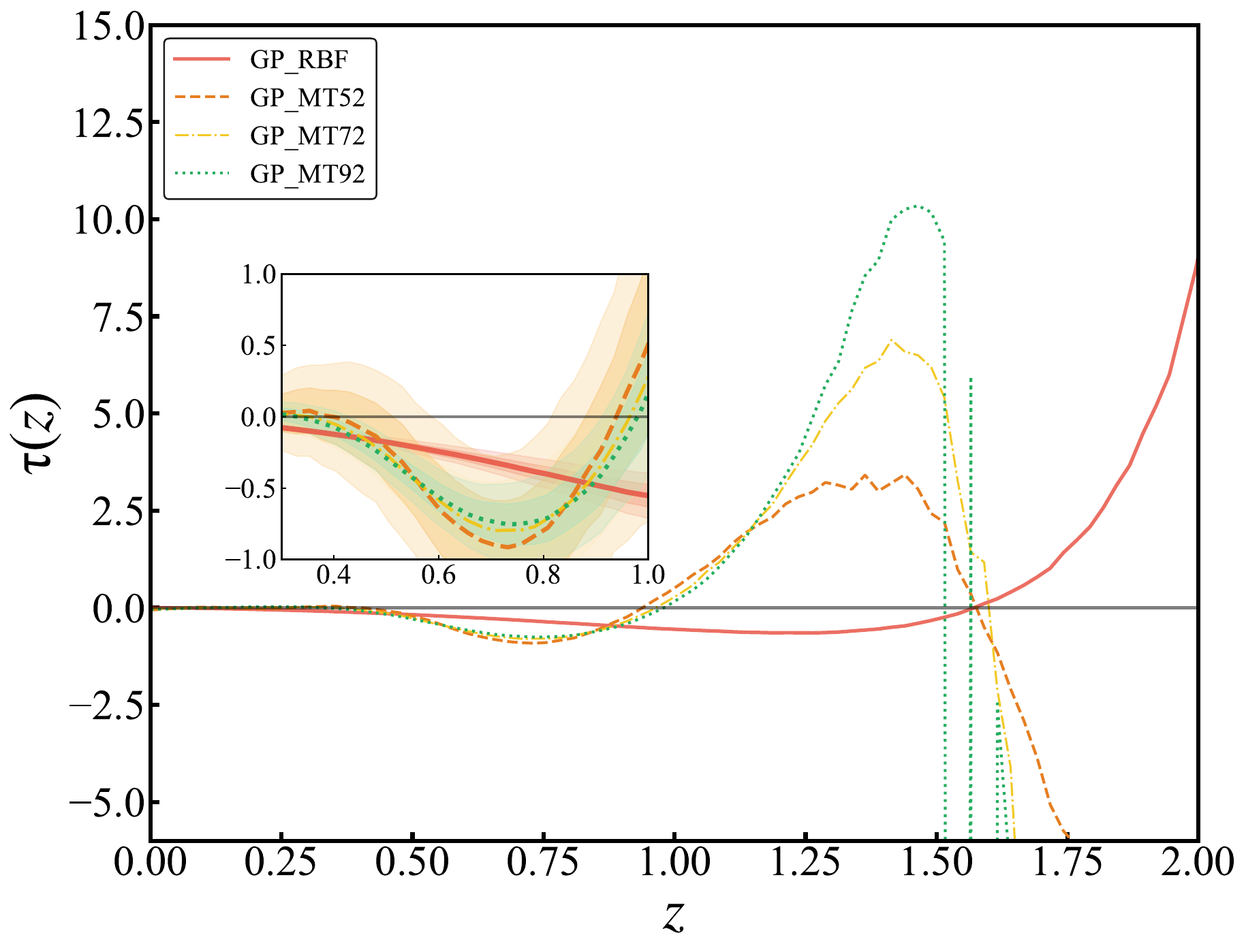}
        \caption{$\tau(z)$ reconstruction (Planck 2018 prior)}
        \label{fig:tau_planck}
    \end{subfigure}

    \vspace{0.5cm}

    \begin{subfigure}[t]{0.48\textwidth}
        \centering
        \includegraphics[width=\textwidth]{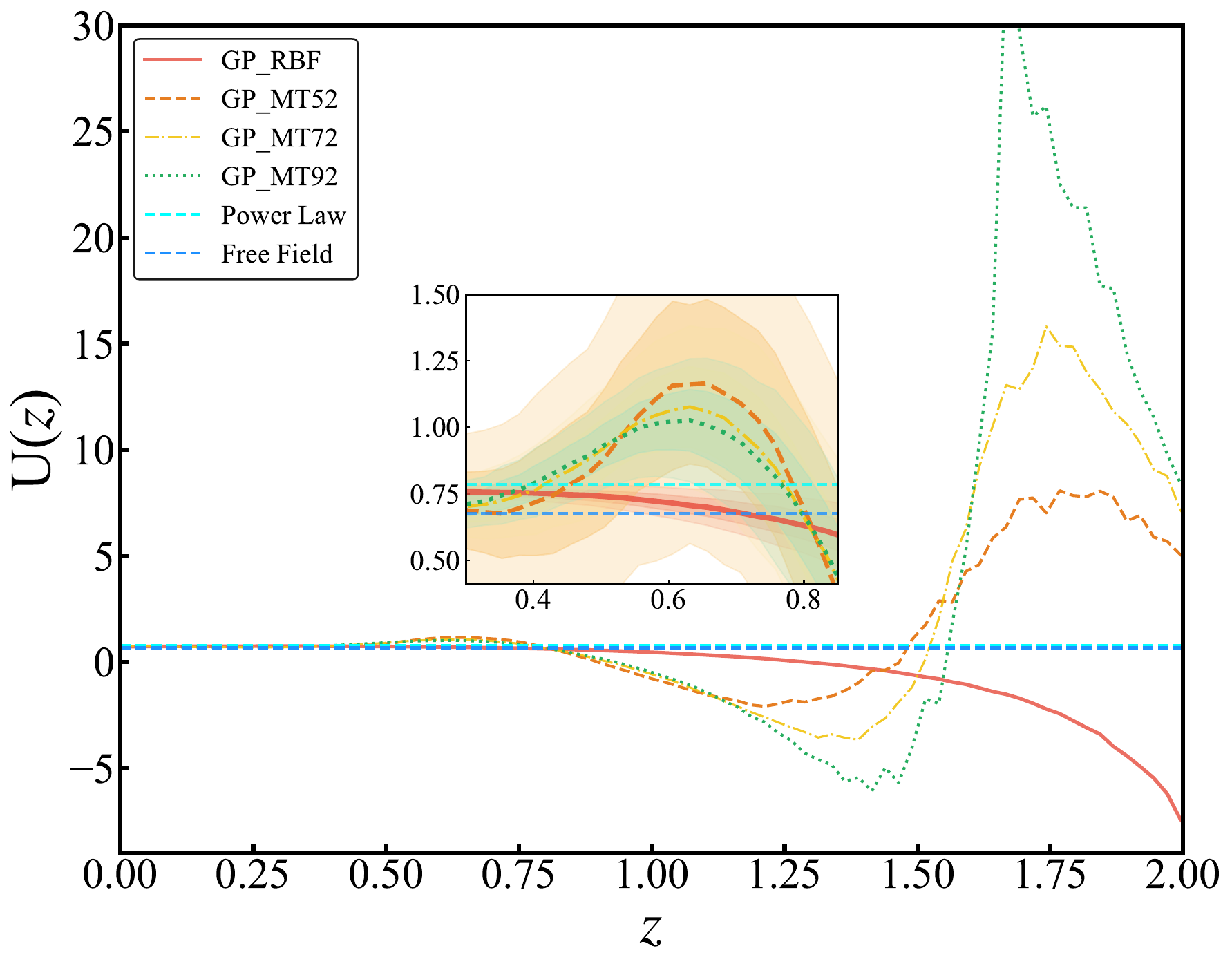}
        \caption{$U(z)$ reconstruction (PantheonPlus+SH0ES prior)}
        \label{fig:u_pantheon}
    \end{subfigure}
    \hfill
    \begin{subfigure}[t]{0.48\textwidth}
        \centering
        \includegraphics[width=\linewidth]{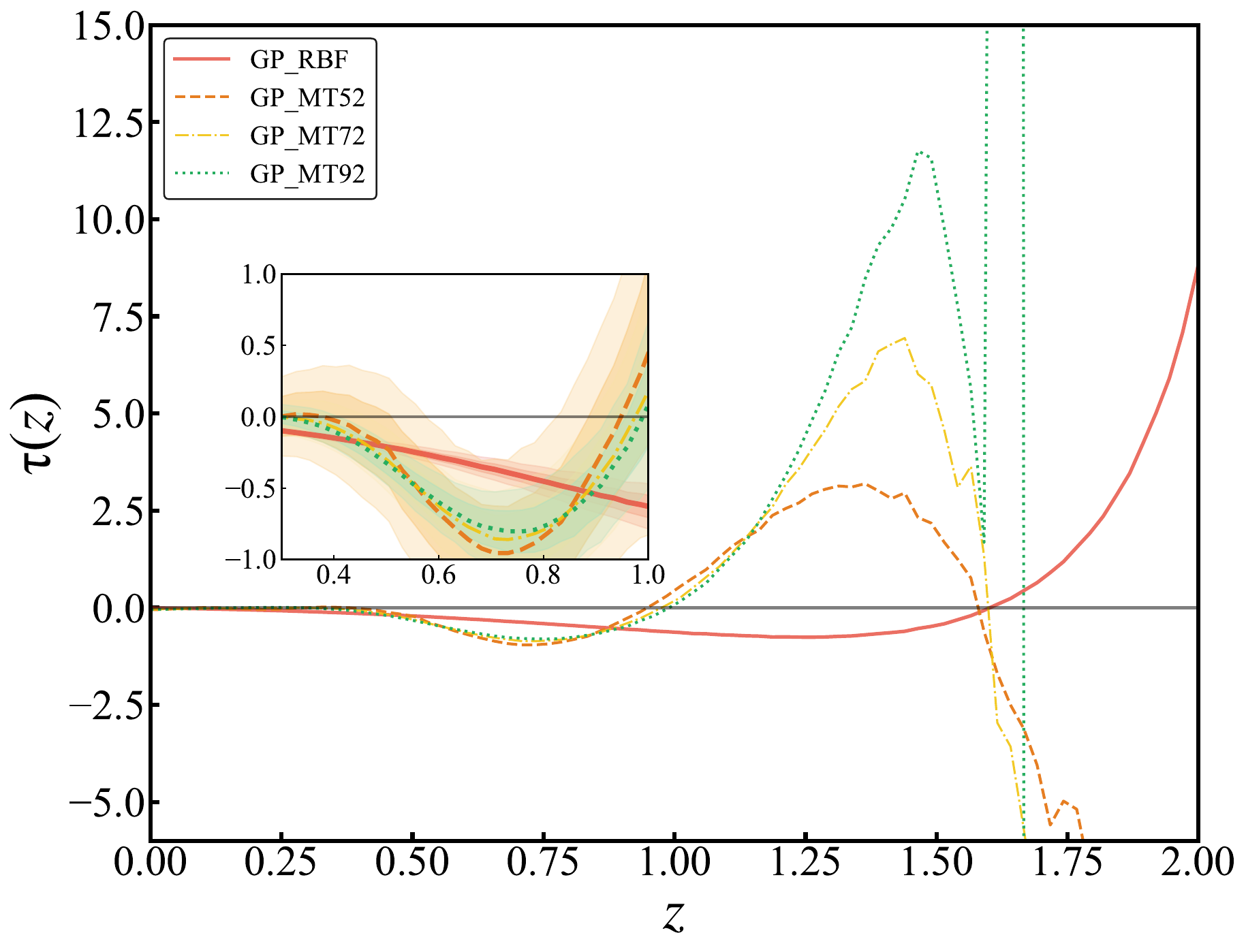}
        \caption{$\tau(z)$ reconstruction (PantheonPlus+SH0ES prior)}
        \label{fig:tau_pantheon}
    \end{subfigure}

    \caption{Comparison of $U(z)$ and $\tau(z)$ reconstructions using different kernel functions under two distinct priors. \textbf{Top row}: Results with the Planck~2018 prior. \textbf{Bottom row}: Results with the PantheonPlus+SH0ES prior.}  In all panels, the inner and outer contours correspond to the $1\sigma$ and $2\sigma$ confidence regions.
    \label{fig:reconstruction_comparison}
\end{figure*}

The reconstruction of $U(z)$ and $\tau(z)$ requires priors on $\Omega_{m0}$ and $H_0$, and we employ the two distinct priors described in Section \ref{datamet} to ensure robustness. Notably, the reconstructed $U(z)$ and $\tau(z)$ exhibit minimal dependence on these priors, attributed to our dimensionless formulation where the primary difference between priors lies in $\Omega_{m0}$, and Eqs.~\eqref{UzDC_flat} and \eqref{tau_final} show weak sensitivity to $\Omega_{m0}$ variations within observational uncertainties ($\Delta\Omega_{m0}\approx0.02$). This prior-insensitivity is visually evident in Fig. \ref{fig:reconstruction_comparison}, where the reconstructed $U(z)$ and $\tau(z)$ bands under the Planck and SH0ES priors show substantial overlap over the entire redshift range.

By sampling these priors alongside the multivariate Gaussian distributions of $D_M(z)$ and its derivatives, we obtain the reconstruction results shown in Fig. \ref{fig:reconstruction_comparison}. The left panels display $U(z)$ with 1$\sigma$ and 2$\sigma$ uncertainty regions, compared with $U_{\mathrm{PL}}$ and $U_{\mathrm{FF}}$, while the right panels present $\tau(z)$, with all panels incorporating results from four GP kernels.

\subsection{Scalar Field Potential Reconstruction}

The reconstructed $U(z)$ from both priors shows distinct constraint characteristics across redshift ranges, as shown in the left panels of Fig.~\ref{fig:reconstruction_comparison}. At low redshifts ($z < 1$), abundant data and minimal uncertainties lead to strong constraints, reflected in narrow 2$\sigma$ confidence intervals with well-defined boundaries, indicating high reconstruction reliability. At high redshifts ($z > 1$), constraining power diminishes, with expanding 1$\sigma$ intervals due to scarce data and increased measurement uncertainties. In the intermediate range ($0.5 < z < 0.85$), the RBF kernel produces nearly fluctuation-free reconstructions with superior noise suppression, while Matérn-class kernels exhibit heightened sensitivity to individual data points, leading to increased oscillations and broader confidence intervals.

All four kernels reveal a consistent trend of decreasing $U(z)$ with increasing redshift for $z<1$, indicative of dark energy's repulsive force evolving to dominate over gravitational attraction. This aligns with the $\Lambda$CDM "low-to-intermediate redshift phase transition," marking the onset of cosmic acceleration.

The reconstruction exhibits minimal dependence on the choice between the \textit{Planck} and SH0ES priors. This is because the primary difference in our dimensionless formulation is $\Omega_{m0}$, and the terms containing $\Omega_{m0}$ in Eq.~\eqref{UzDC_flat} are subdominant to those involving $D(z)$ and its derivatives. The uncertainty in $\Omega_{m0}$ is also small relative to the total reconstruction uncertainty.

The power-law potential $U_{\mathrm{PL}}$ overlaps with the $1\sigma$ interval of the reconstructed $U(z)$ at low redshifts ($z < 0.3$), lies between the $1\sigma$ and $2\sigma$ bands at intermediate redshifts ($0.3 < z < 1.7$), and re-enters the $1\sigma$ interval at high redshifts ($z > 1.7$). This redshift-dependent behavior indicates that $U_{\mathrm{PL}}$ remains compatible with current data constraints in both the early and late universe, while showing weaker consistency during the transition epoch.

It is important to note that the power-law and free-field potentials are presented here only for contextualizing the model-independent $U(z)$ reconstruction results, not for formal statistical model comparison-a rigorous model comparison would require quantitative statistical tests  that are beyond the scope of this work's non-parametric reconstruction framework. For the sake of objective comparison, we also note that the power-law potential $U_{\text{PL}}(\phi)$ deviates beyond the $2\sigma$ uncertainty interval of the reconstructed $U(z)$ for all kernels at the intermediate redshift range $0.9 < z < 1.1$, indicating that neither canonical scalar field potential provides a fully consistent description of dark energy evolution across the entire redshift range probed in this work.

The quadratic free-field potential $U_{\text{FF}}(\phi)$ falls within the $1\sigma$ uncertainty interval for approximately 60\% of the redshift range across all four GP kernels. However, it exhibits a significant reduction in overlap and deviates beyond the $2\sigma$ uncertainty interval for all Matérn-class kernels at intermediate redshifts ($0.5 < z < 0.8$), particularly under the Planck 2018 prior; the RBF kernel, by contrast, remains marginally consistent with $U_{\text{FF}}(\phi)$ at the $2\sigma$ level in this redshift range. At high redshifts ($z > 0.8$), the consistency between the reconstructed $U(z)$ and $U_{\text{FF}}(\phi)$ decreases markedly across all four kernels, with the reconstructed results deviating beyond the $2\sigma$ uncertainty interval. In this high-redshift region, the RBF kernel exhibits a relatively smooth variation, while the Matérn-class kernels show more pronounced fluctuations. This collective behavior across the majority of kernels suggests that current observational data provide only limited support for $U_{\text{FF}}(\phi)$ in describing the evolution of dark energy in the late universe.

Among the four kernels, the RBF kernel produces the narrowest uncertainty bands at $z < 1$ due to its superior noise suppression. This characteristic helps quantify the range of potential forms consistent with observational data, rather than enabling definitive discrimination between specific quintessence models.

\subsection{Kinetic Energy Reconstruction}

The right panels of Fig. \ref{fig:reconstruction_comparison} illustrate the reconstructed $\tau(z)$ functions obtained with different kernels. At low redshifts ($z<0.5$), the results are consistent and $\tau(z)$ approaches zero. Around $z \approx 1.0$, $\tau(z)$ crosses zero: the Matérn-5/2 kernel gives the sharpest crossing, while the RBF kernel exhibits a notably gradual transition. This gradual trend, caused by the RBF kernel’s strong smoothness and noise suppression, completes its zero-crossing around $z \approx 1.6$-a delayed appearance compared to other kernels. Importantly, this delay is a methodological artifact of the RBF kernel, which predicts a delayed zero-crossing at $z \sim 1.2$. However, we attribute this feature to a methodological artifact of the kernel rather than to a physical signal beyond $\Lambda$CDM. First, such a delayed transition would imply an unphysically late onset of cosmic acceleration, inconsistent with independent constraints from cosmic chronometers and the BAO scale at $z \sim 1.47$, all of which favor an earlier transition epoch. Second, this behavior arises from the RBF kernel's inherent infinite smoothness, which imposes a strong regularization prior that artificially suppresses rapid variations at high redshifts ($z > 1.5$), rigidly delaying the transition. In contrast, Mat\'ern kernels (M52/M72/M92) allow for a degree of roughness more consistent with the noise structure of the data. Third, at $z > 1.5$, the reconstruction becomes prior-dominated rather than data-driven, as the number of Pantheon+ supernovae and DESI BAO data points decreases sharply beyond this redshift, leaving the RBF kernel's smooth extrapolation largely unconstrained by empirical evidence. For these reasons, we interpret the delayed zero-crossing and high-redshift behavior of the RBF kernel not as genuine features of dark energy evolution, but as consequences of its regularization assumptions.

Physically, $\tau(z)$ evolution reflects the scalar field's dynamical properties, with the zero-crossing at $z \approx 1.0$ corresponding to the cosmic equality epoch where $\Omega_\phi = \Omega_m$, marking the onset of dark energy domination and cosmic acceleration, consistent with standard expectations.

An intriguing feature in the mean reconstruction is the emergence of negative values for $\tau(z)$ in the intermediate redshift range ($0.5<z<1.0$), which at face value conflicts with the theoretical expectation $\tau(z) \geq 0$ for a canonical scalar field. However, as noted earlier, these values are consistent with zero within the $1\sigma$ uncertainties. Their appearance can be attributed to two factors: first, error amplification, $\tau(z)$ depends on the second derivative of $D_M(z)$ (Eq. \ref{tau_final}), which amplifies measurement uncertainties and reconstruction errors; second, reduced data coherence.

Notably, the sources of spurious behavior identified in this work are not limited to negative $\tau(z)$ at intermediate redshifts, but have been effectively mitigated through targeted strategies that enhance the reliability of our results. First, the combined DESI DR2 BAO and Pantheon+SH0ES SN Ia datasets provide high-density coverage at low to intermediate redshifts ($z<1$), which significantly suppresses error amplification. Second, cross-validation using four GP kernels allows us to robustly distinguish physical signals from kernel-dependent artifacts, with the RBF kernel's superior noise suppression further filtering out unphysical fluctuations. These advantages yield consistent trends across all kernels in the core redshift range ($z<0.7$), validating the robustness of our reconstruction.

Meanwhile, the underlying sources of artifacts continue to influence reconstructions across the full redshift range: at $z>1.5$, weakened data constraints lead to broadened uncertainty bands for all kernels, and fine-scale features may include spurious fluctuations; oscillations in $U(z)$ from Matérn-class kernels have also been identified as artifacts through multi-kernel comparison. It is important to emphasize that our nonparametric framework does not presuppose any functional form for the potential or kinetic energy, thereby avoiding systematic biases that could be introduced by parametric models. This allows the data to drive the extraction of physical signals, a feature particularly valuable for artifact discrimination. Even in the presence of local artifacts, core physical trends, such as the zero-crossing of $\tau(z)$ and the decreasing behavior of $U(z)$, are consistently confirmed through multi-kernel agreement and data density analysis.

\section{\label{conc}Conclusions}

This study presents a model-independent framework for reconstructing the quintessence scalar field potential and kinetic energy, leveraging DESI DR2 BAO and PantheonPlus+SH0ES SN Ia data. By employing GP  with four covariance kernels (RBF, Matérn-5/2, Matérn-7/2, Matérn-9/2), we avoid parametric model biases, enabling direct extraction of dark energy dynamics from data without functional form assumptions.

Our key findings are summarized as follows: At low redshifts ($z<0.7$), reconstructions across all kernels yield consistent results with overlapping $1\sigma$ intervals, confirming the reliability of our nonparametric approach and the moderate constraining power of the combined dataset in this range. The RBF kernel produces narrower uncertainty bands than Matérn-class kernels (particularly at $z\approx1$) and effectively captures smooth trends due to its noise suppression capability, making it a useful tool for cosmological reconstruction. The power-law potential $U_{\mathrm{PL}}$ aligns with reconstructed $U(z)$ within $1\sigma$ for $\sim 70\%$ of the redshift range (with good overlap at $z<0.3$ and $z>1.5$), indicating it is one of the viable dark energy models compatible with current data; in contrast, $U_{\mathrm{FF}}$ shows reduced overlap with data constraints at intermediate redshifts ($0.5<z<0.8$), reflecting weaker support from observations. The reconstructed $\tau(z)$ exhibits non-monotonic evolution, increasing to $z\approx1.0$ and then decreasing, with a zero-crossing corresponding to the cosmic equality epoch and the onset of dark energy domination. Apparent negative $\tau(z)$ values in the intermediate range are statistical artifacts arising from derivative error amplification and reduced data coherence, and remain consistent with zero within $1\sigma$ uncertainties.

Our findings emphasize the need for improved observations at intermediate and high redshifts ($z>1.0$) and highlight key strengths of this work. The combined DESI DR2 BAO and Pantheon+SH0ES SN Ia dataset spans $0.01<z<2.3$, offering dense low‑to‑intermediate redshift constraints that reduce error amplification from higher‑order derivatives. By comparing four GP kernels of different smoothness, we cross‑validate results to separate physical signals from artifacts. The dimensionless formulation minimizes dependence on priors such as $\Omega_{m0}$, preserving physical meaning for $U(z)$ and $\tau(z)$ even where constraints weaken. However, for $z>1.5$ confidence intervals broaden significantly, and both $U(z)$ and $\tau(z)$ become prone to kernel‑dependent oscillations and noise‑amplified fluctuations beyond the intermediate‑redshift negative $\tau(z)$ feature, revealing current limitations in constraining early universe dark energy dynamics. Future datasets with more SN Ia and BAO measurements at these redshifts will refine scalar field reconstructions and test quintessence validity.

By establishing a robust, data-driven approach to probe dark energy, this work narrows the gap between observational constraints and theoretical models, paving the way for deeper investigations into cosmic acceleration and new physics beyond $\Lambda$CDM. Future extensions will incorporate additional probes (e.g., cosmic chronometers, weak lensing) and advanced kernel functions to improve reconstruction precision and physical interpretability.

\section*{Acknowledgments}
We are grateful to the anonymous reviewer for their professional suggestions, which have significantly improved the quality of the manuscript.
This work was supported by the National Natural Science Foundation of China under Grant No. 12203009; the National Key Research and Development Program of China (No. 2024YFC2207400); and the Chutian Scholars Program in Hubei Province (X2023007).

\bibliographystyle{apsrev}
\bibliography{references}

\end{document}